\documentclass[letterpaper,11pt]{article}

\raggedright
\usepackage[svgnames]{xcolor}
\usepackage{framed}
\usepackage{tocloft}
\usepackage{color, soul}
\usepackage{authblk}
\usepackage{graphicx}
\usepackage[latin1]{inputenc}
\usepackage[T1]{fontenc}
\usepackage{times}
\usepackage[toc,page]{appendix}
\usepackage{natbib}


\author[1]{Yimin Luo}
\author[1]{Francesca Serra}
\author[1,*]{Kathleen J. Stebe}
\affil[1]{Department of Chemical and Biomolecular Engineering, University of Pennsylvania, 220 South 33rd Street, 311A Towne Building, Philadelphia, PA 19104}
\begin{document}

\title{Experimental realization of the \\``lock-and-key'' mechanism in liquid crystals}
\date{}
\maketitle
\begin{abstract}

The ability to control the movement and assembly of particles in liquid crystals is not only  an important route to design functional materials, but also sheds light on the mechanisms of colloidal interactions. In this study we place micron-sized particles with ``Saturn ring'' defects near a wall with hills and dales that impose perpendicular (homeotropic) molecular anchoring. The strong splay distortion at the wall interacts with the distortion around the particles in the near field and favors their migration towards the dales via the so-called ``lock-and-key'' mechanism. We demonstrate experimentally that the lock-and-key mechanism can robustly localize a particle at specific topographical features. We observe the complex trajectories traced by the particles as they dock on the dales, estimate the binding energy, and explore a range of parameters that favor or disfavor the docking event, thus exploiting the capabilities of our experimental system. We extend the study to colloids with homeotropic anchoring but with an associated point defect instead of a Saturn ring and show that they find a different preferred location, i.e. we can place otherwise identical particles at well defined sites according to their topological defect structure.  Finally, for deep enough wells, confinement drives topological transitions of Saturn rings to dipoles.  This ability to tailor wall geometry to guide colloids to well defined sites within nematic liquid crystals represents an important new tool in directed assembly.
\end{abstract}

~\\

Fabrication through directed self-assembly provides a major simplification in processing, creating, and controlling supramolecular materials.  
This approach has revolutionized polymer science and is a modality that pervades biology.\cite{whitesides2002self, manoharan2015colloidal} Soft materials are ideal  systems to create long-range interactions by molding the associated free energy fields within which inclusions move, interact and assemble. For example, by molding the shape of isotropic fluid interfaces, anisotropic particles assemble via capillary forces.\cite{cavallaro2011curvature}  Nematic liquid crystals (NLCs) are excellent candidates for such an approach; when they are confined within boundaries that prevent a uniform director orientation, topological defects can ensue -- small ``melted'' regions (points or lines) where all the disorder is concentrated. The location of these defects can be organized by confinement: in  recent literature, micro- and nano-particles have been successfully assembled by point and line defects.\cite{blanc2013ordering,hegman2007,pires2007colloid} In those systems, the inclusions were drawn to the melted cores irrespective of their surface chemistry, orientation, or geometry, and were simply sequestered in the higher energy, higher symmetry core. The colloidal inclusions can, however, be treated to induce surface anchoring themselves. In that case it is possible to refine their response based on the local geometry of the director field. This is most striking when the colloids induce topological defects of their own, allowing assembly to occur without the ``middleman'' of bulk defects. \cite{muvsevivc2006two} The defect structures are now {\sl mobile} and move into place with their instigating inclusion.
If, for instance, a particle has planar (tangent to the surface) anchoring, it forms two defects (boojums) at two poles of the particle. On the other hand, if a particle has homeotropic anchoring, either a disclination loop forms around it (a Saturn ring), producing a elastic quadrupole, or a point defect forms nearby, creating a elastic dipole. A mechanism to sort particles on the basis of the associated defect was proposed by Nikkhou {\it et al.},\cite{nikkhou2015light} in which defect loops around a thin fiber can attract particles with homeotropic anchoring and are able to differentiate the orientation of the elastic dipoles.

~\\

Telo de Gama and coworkers proposed another  possible motif  for particle sorting in a {\sl bulk defect free} NLC: the ``lock-and-key'' mechanism,\cite{silvestre2004key} inspired by complementary protein binding domains in biology and exploited in other contexts in colloidal science. \cite{sacanna2010lock, manoharan2015colloidal} Two-dimensional (2D) simulation predicted that a homeotropic disk in a NLC minimizes its energy by nestling into a well also with homeotropic anchoring and with the same radius of curvature as the particle. Full three-dimensional studies confirmed the initial concept,\cite{eskandari2014particle} and suggested a mechanism of sorting the particles between topographical peaks and wells, or, as Maxwell would call them, ``hills and dales'', based on the anchoring of the particles. The simple rule is that particles migrate to the dales if the anchoring of the particle and the substrate coincide (i.e. both homeotropic or both planar) and to the hills if the anchoring is different for particle and substrate. We should remark that in the three-dimensional simulations the homeotropic particles always have associated Saturn ring defects. Simulation also highlighted the importance of the geometry of the cavity.  \cite{hung2007nanoparticles} This phenomenon was also explored in the work by Silvestre {\it et al.} \cite{silvestre2014towards} that showed experimentally higher probability for particles with planar anchoring to locate at tip of a homeotropic micro-pyramid, and for those with homeotropic anchoring to locate at the base of the same pyramid.

~\\

Here we present an experimental system that confirms the lock-and-key mechanism for Saturn ring particles, and that also allows us to interrogate the  behavior of the elastic dipoles. To demonstrate the lock-and-key mechanism, we designed an experimental system (i) to have a 2D-like geometry as in the original simulation,\cite{silvestre2004key} (ii) to control independently the anchoring at the wall and at the particles, (iii) to allow for clear visualization of the process of a particle moving towards topographical features and finding its preferred location, (vi) to  investigate particles with Saturn rings but also particles with an associated point defect, and (v) to potentially allow for the exploration of many parameters and anchoring configurations.

~\\

Our docking sites are periodic indentations and protrusions in a long wavy wall.  The wall has length $1.5$ cm, width 150 $\mu$m and thickness 20 to 25 $\mu$m. The indentations and protrusions have width $w$=20 $\mu$m, radius of curvature $R$=7.5 $\mu$m and the amplitude of the wavy structure is $h$=5.0 $\mu$m (Fig. 1(a)). The walls were fabricated by photolithography with SU-8 2015 (MicroChem Corp.) on a glass cover slip and treated to have strong homeotropic anchoring using DMOAP (Dimethyloctadecyl[3-(trimethoxysilyl)propyl] (Sigma-Aldrich) after being coated with silica by chemical vapor deposition. 
Once formed, the wall was then carefully removed from the substrate with a razor blade to form a strip: in this way we could obtain a floating strip in which three sides, including the ``wavy'' one, had homeotropic anchoring, and one (the one that was previously in contact with the glass) retained the native degenerate planar anchoring from SU-8. Two glass cover slips were prepared with uniform planar anchoring (coated with 1 wt\% poly(vinyl alcohol) in DI water, annealed for an hour at 80$^\circ$C and rubbed with a velvet cloth). The treated strip was then carefully placed between the cover slips to form a planar cell ``sandwiching'' a wavy wall. The cover slips were aligned in an anti-parallel fashion to minimize splay due to pre-tilt angle in the cell, which destablizes Saturn rings even under confinement and creates a bias in the direction of the dipoles. The strips were arranged with their longest axes perpendicular to the direction of the planar alignment, as shown in Fig. 1(b). We secured the cover slips with binder clips. The pressure ensured that there would be no gap between the strips and the glass, so the side of the strip with planar anchoring would not introduce bulk defects. In this way, the sides of the strips in contact with the LCs were the straight homeotropic wall and the wavy homeotropic wall. Interactions of colloids with this wavy wall are the focus of this study.

~\\

\begin{figure}[h]
\centering
  \includegraphics[height=8cm]{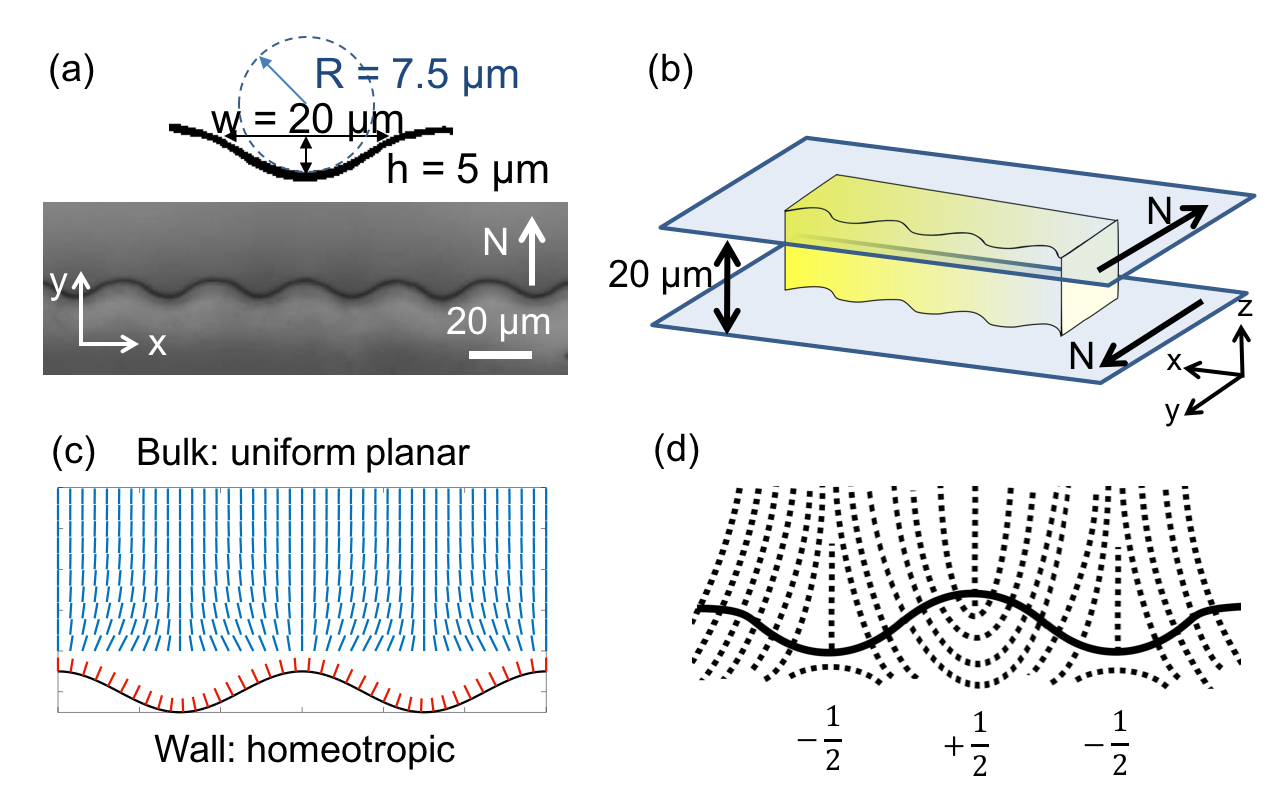}
  \caption{(a) Microscope image of the SU8 structure (top view) with the description of the relevant geometrical parameters for the hills and dales (inset). (b) Scheme of the assembled cell. The vector N indicates the direction of rubbing. (c) Simulated director field in the small-slope limit.  (d) Field created by alternating point defects with half integer charge embedded in the wall resembles the field created by the hills and dales. }
  \label{fig1}
\end{figure}

~\\

We loaded a well-dispersed suspension of silica particles (2$a$ = 5 or 15 $\mu$m, 1 wt\%, Corpuscular Inc., where $a$ denotes particle radius), treated for homeotropic anchoring, in 5CB (4-Cyano-4'-pentylbiphenyl, Kingston Chemicals). 5CB in the isotropic phase is inserted into the assembled cell via capillary action and subsequently cooled below the clearing point ($T_{NI} \approx$ 34.9 $^\circ$C). 
The cell was imaged with an upright microscope (Zeiss AxioImager M1m) in transmission mode equipped with crossed polarizers and the images were captured by a high-resolution camera (Zeiss AxioCam HRc) under magnification ranging from 20x to 50x. Before observing the colloidal dynamics in the cell, we confirmed the good planar alignment of the glass coverslips and the good homeotropic anchoring on the wall by polarized optical microscopy. 

~\\

As a result of their curvature, the gentle hills and dales of the walls become sites of distortion in the nematic director field $\bf n$. The distortion in the bulk decays smoothly away from the wall without introducing any topological defects. It is instructive to consider a Fourier mode of the function defining the wall shape:

\begin{equation}
    h(x) = A cos(kx)
 \end{equation}
 
\noindent under the single elastic constant approximation and assuming small gradients in the director field. In this limit,  $\bf n$ can be expressed simply (see  Fig. 1(c)): 

\begin{equation}
    \left\{
                \begin{array}{ll}
                  n_x = Aksin(kx)e^{-ky}\\
                  n_y= 1
                \end{array}
              \right.
 \end{equation}
 
~\\

\noindent where A is the amplitude of the sinusoidal wave, k is the wave number, h(x) is the position of the wavy wall, and $n_x, n_y$ are the components of the director \textbf{n} along $\hat x$ and $\hat y$. This director field could be realized either near walls with shallow hills and dales, or, alternatively, in regions far enough from the wall that gradients in $\bf n$ are weak. This form suggests that hills and dales are sites with the highest splay energy density, while the inflection points are sites with highest bend energy density. Furthermore, it indicates that distortions decay over distances comparable to the wavelength of the features, limiting the range of interaction with colloids in the gap. Closer to the wall, however, the field created by the hills and dales resembles a field created by an array of point defects embedded in the wall with semi-integer charge, with alternating signs, as shown schematically in Fig. 1(d). The distortion created by a particle with homeotropic anchoring always induces a defect, either an elastic dipole or a Saturn ring.\cite{terentjev1995disclination} To investigate the lock-and-key mechanism predicted in  simulation,\cite{silvestre2004key, eskandari2014particle} it was essential to have particles with Saturn rings. We achieved this by utilizing large silica particles ($2a$ = 15 $\mu$m) whose size not only matched the size of the ``lock'', but also was only slightly smaller than the total thickness of the LC cell (20-25 $\mu$m). Confinement from the walls of a LC cell is known to stabilize Saturn rings against point defects.\cite{stark} In particular, the Saturn ring becomes the most stable configuration when the ratio between the radius of the particle and the distance between the particle surface and the wall is bigger than unity. In our case, given the polydispersity of the particles and the uncertainty in the cell thickness, this ratio was between 0.9 and 2.5.  

~\\

Flat walls with homeotropic anchoring are repulsive to homeotropic colloids \cite{lavrentovich2014transport}, as we have verified also in our system. Here, however, wall curvature and the associated splay and bend energies guide the colloid to its preferred site. This occurs only for slowly moving particles located sufficiently close to the wall, \textit{i.e.} for distance less than 5 $\mu$m, so the elastic energies remain pronounced, and for average drift velocities less than roughly 5 $\mu$m s$^{-1}$, so distortions of the director field associated with viscous flow are weak. Provided these conditions are obeyed, however, docking is remarkably robust. Of 29 observations, 14 did not obey these conditions, i.e. they were too far away and/or moving too quickly to dock ({\it e.g.} ESI Video 1.).  The remaining 15 particles did obey these criteria; all found their docking sites. Trajectories, observed in full for 5 cases and in part for 2 others, reveal a complex energy landscape. Close to the edge of the structured wall, a colloidal particle is attracted towards the dale as shown in Fig. 2(a-e) and ESI Video 2. The particle, initially situated near a hill on the wavy wall, moves away from this position due to the unfavorable distortion of the director field (Fig. 2(a)). As it floats past the hill, the Saturn ring is visibly distorted away from the equatorial plane to best accommodate the distortion of the external field (denoted by white arrows), indicating that the distortion around the colloidal particle is interacting with the distortion near the wall (Fig. 2(b)). As the particle makes its way toward the dale, the ring is once again distorted, this time lifting a bit, indicating the change of local director field (Fig. 2(c)). Finally, the particle slowly finds its equilibrium position as it sits comfortably in the dale, at the minimum of the shallow well in the wall. As this occurs, the Saturn ring remains distorted (Fig. 2(e)). Here the lock-and-key mechanism appears clearly. In Fig. 2(f), we present a schematic of the director field around a colloid with a distorted Saturn ring defect docked in a dale. The director field around the colloid matches very well with the director field inside the dale.

~\\

\begin{figure}
 \centering
 \includegraphics[height=6cm]{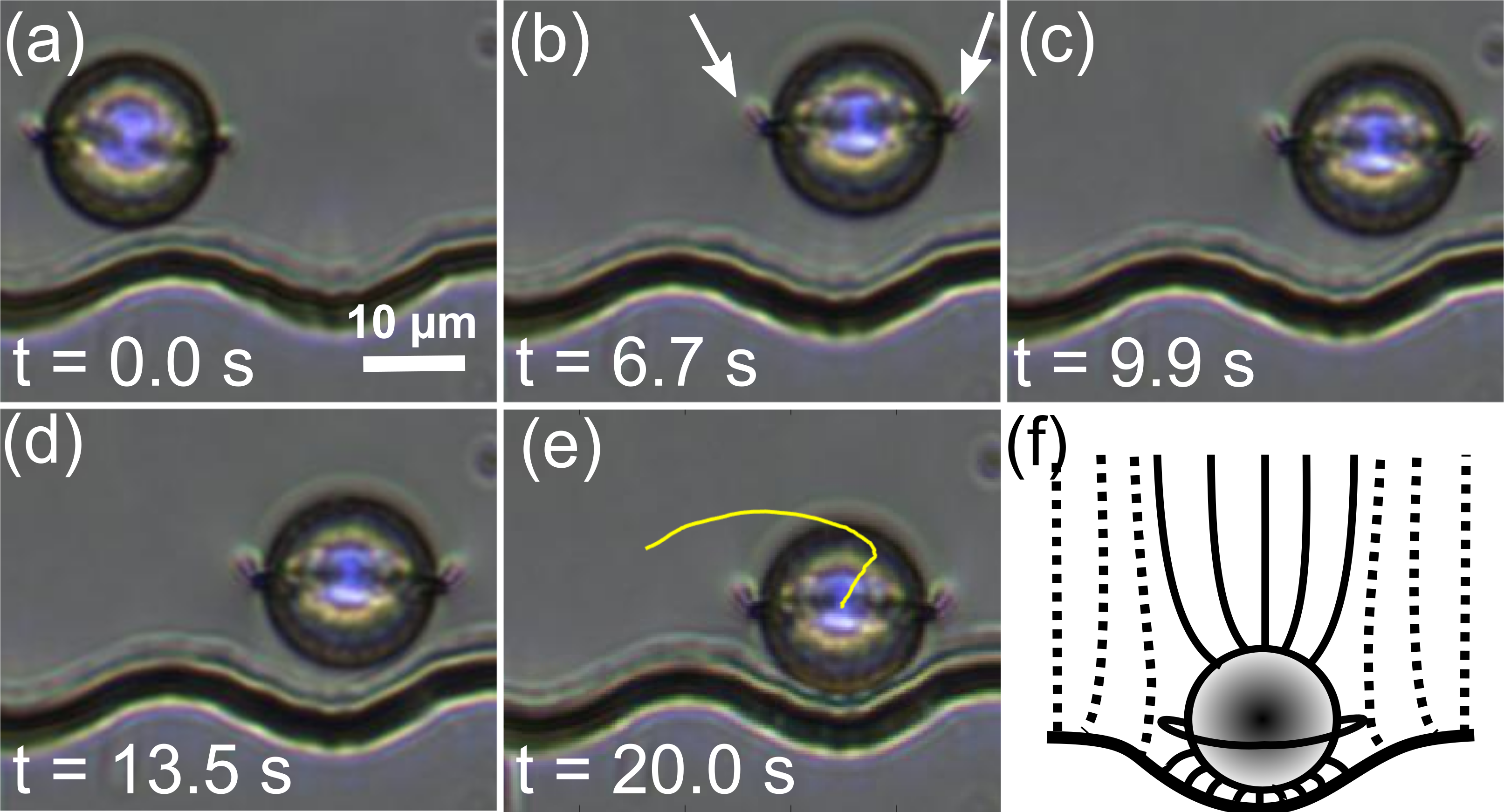}
 \caption{(a-e) Time sequence of the particle ($2a$ = 15 $\mu$m) moving into a cavity of comparable size. White arrows in (b) denote distortion of the Saturn ring. Yellow line in (e) denotes the trajectory of the movement. (f) Schematic of the director field once the particle docks on the dale, minimizing the total splay distortion.}
 \label{fig2}
\end{figure}

The trajectory traced by the particle's center of mass is shown in Fig. 3(a); the bold dots demarcate distances of 5 $\mu$m in arclength $s$ from the initial position of the particle center of mass, which defines $s=0$. We have observed the signature features of these complex interactions in $5$ separate experiments. In the initial stage of the migration, the particle slows down (Fig. 3(b)), owing to viscous dissipation and the elastic energy field near the wall. As the particle migrates past the hill, it accelerates in both $\hat x$ and $\hat y$ directions, (Fig. 3(b)), as it is ``kicked'' up diagonally  (point \textbf{A} in Fig. 3(a)).  The particle then moves toward the dale, and slowly overshoots this location (point \textbf{B}). At this point, it reverses its direction. Thereafter, it slows down significantly in the last portion as it finds its way into the final docking position (point \textbf{C}).  Since the kick-reversal-docking behavior is reproducible, we define the binding energy to have contribution from all three components, {\it i.e.} the entire trajectory as presented in Fig 3(a).

~\\

We measure the average velocity of the particle, and determine the Reynolds number $Re$ = 3 $\times 10^{-6}$ and the Erickson numbers $Er=0.1$ using typical values for the material constants (see appendix). In this limit, viscous forces balance elastic forces that drive the particle's motion, and the elastic energy landscape is not perturbed by the particle's motion. This allows the binding energy to be estimated from the energy dissipated over the particle's trajectory. Since the particle moves in a thin gap of half thickness $l \sim a$ and within a distance $d \sim a$ to the wavy wall (Fig. 3(c)), significant corrections to the viscous drag on the sphere must be addressed to account for these near-wall hydrodynamic interactions. To our knowledge, there are neither analytical results nor simulations of spheres in Stokes flow in a setting akin to our experiment. However, Ganatos et al. \cite{ganatos1980strong} provide exact solutions for drag on a sphere moving in a thin gap and O'Neill provides drag coefficients for particles parallel to a wall.\cite{o1964slow} We estimate the effective drag coefficient by adding these contributions; while this approximation neglects details of the flow in the gap interacting with the flow near the wall, it suffices for an order of magnitude estimate of the interaction energy. The drag coefficient $C_D$ estimated in this way is roughly $4.0$ (see appendix).
In the NLC 5CB, the viscosity is anisotropic, differing by a factor of roughly 1.6 for motion transverse to or along the director.  As most of the particle migration occurs in the $x$ direction, we adopt the value for the viscosity in the transverse direction, {\it i.e.} $\eta$ =14.3 mPa s, as inferred from the diffusivity measurements of 5 $\mu$m-sized particles in 5CB reviewed by Lavrentovich.\cite{lavrentovich2014transport} To estimate the energy dissipation U from the initial to final positions, we integrate along the particle path:

\begin{equation}
U = {C_D}6\pi \eta a\int\limits_0^{{s_f}} {v} d{s}
\end{equation}

\noindent where $C_D \approx$ 4 is the effective drag coefficient for this setting, $v$ is the velocity of the particle tangent to the path, and $ds$ is the arclength element. We find $U \approx 10^5$ $k_BT$, orders of magnitude larger than the usual Van der Waals or electrostatic interactions exploited in typical colloidal assembly schemes. A representative graph of the energy dissipated along the trajectory $U$ vs. $s$ is plotted in Fig. 3(d). This value cannot be compared to simulation, as current computation power precludes detailed simulation of large particles, and addresses energy gradients far weaker than those explored here. Furthermore, most simulations  address colloid-well interactions that occur directly above the well, while here most of the interaction occurs as the particles are repelled from hills and move toward dales. If one were to focus only on the final stages of interaction, during the slow reversal and docking process as the colloid moves from \textbf{B} to \textbf{C} in the last 5 $\mu$m of the trajectory (Fig. 3(d) inset), energy dissipated is $\approx 10^4$ $k_BT$. Lock-and-key interaction of particles with wells on walls have been simulated as depletion-driven interactions for particle-wall gap distances similar to the size of solvent molecules. \cite{konig2009lock} Depletion effects owing to the escape of solvent LC molecules beneath the particle would become significant for gap distances similar to the size of the LC molecules.  As our particles settle to gaps of 2 $\mu$m at the end of the trajectory, we cannot attribute the observed interactions to this effect.

\begin{figure}
 \centering
 \includegraphics[height=9.5cm]{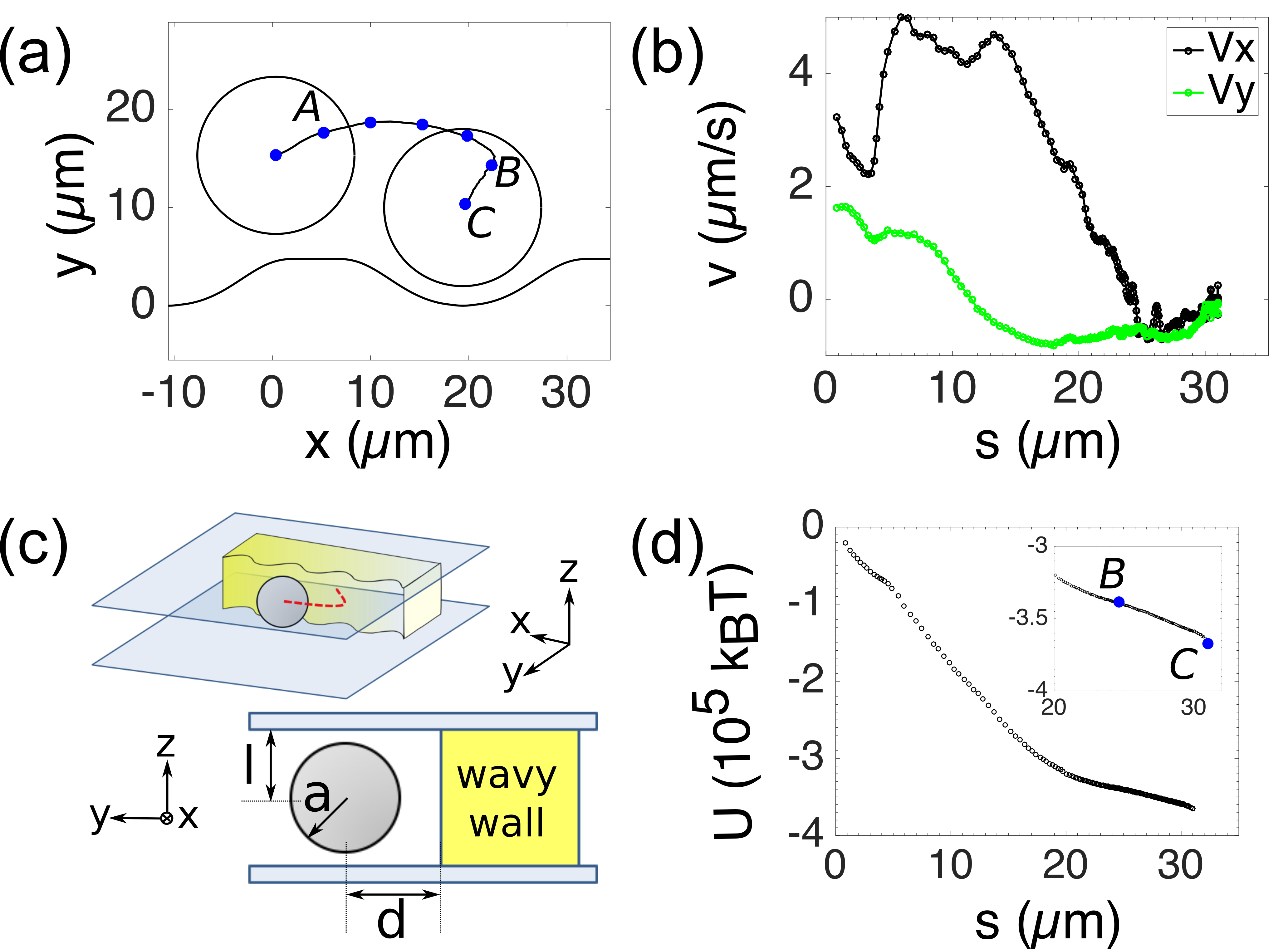}
 \caption{(a) Trajectory of the particle: circles denote the initial and final positions. Bold blue dots denotes increments in arc length $s$ = 0, 5, 10, 15, 20, 25, 30 $\mu$m. Red arrows and capital letters A, B, C denote three events: kick, reverse and docking, respectively. (b) Velocities in $x$- (black) and $y$-directions (green). (c) Particle migrates in a thin gap of half width $l \sim a$ with center of mass a distance $d \sim a$ from wavy wall; hydrodynamic interactions must be addressed in estimate of  $U$. (d) Energy dissipated $U(s)$ over particle trajectory. }
 \label{fig2}
\end{figure}

~\\

Why do the particles nest in the dales? In simulations of quadrupolar defects docking in sharp-edged wells, \cite{silvestre2014towards,silvestre2004key} the ``defect-sharing'' mechanism played a prominent role; the particle merged its defect ring with the region of maximum distortion in the director field near the edge of the well. However, in our case, (Fig. 2), while the Saturn ring is distorted, it is still clearly identifiable. Similarly intact, distorted Saturn rings are shown in Fig. 7 in Hung et al. \cite{hung2007nanoparticles} for particles in equilibrium positions in grooves. This suggests a reduced role for defect sharing, and that splay-matching plays a strong role in the particle docking. The localization of particles is related conceptually to particle sorting in the work by Peng et al., \cite{peng2015liquid} in which splay deformation was created by photo-alignment. The complex, curved trajectory resembles the curved trajectory analyzed by Pires {\it et al.}\cite{pires2007colloid}, and observed by Skarabot {\it et al.}\cite{vskarabot2008hierarchical} and Nikkhou {\it et al.}\cite{nikkhou2015light} for particles docking on a disclination line. In the latter case, the binding energy is very similar to our case, {\it i.e.} U $\approx$ 8000 $k_BT$. As expected, the value we measure for our particles is larger, as our particles are significantly bigger.\cite{vskarabot2008hierarchical} This is remarkable, because the director field in our case is non-singular. However, the similarity of the splay field with that  around a disclination line may explain the curvature of the trajectory.

~\\

The initial position ($s=0$) (Fig. 2a) from which the particle begins its trajectory is roughly one period away from the location where the particle reverses its direction (Fig. 2c). This suggests that particles near this location are metastable, and could execute diverging trajectories, one heading over the hill, the other along the ``reverse'' segment of the trajectory, entering the dale. The distortion of the Saturn ring as the particle travels across different topographical terrains suggests interaction and rearrangement of the director field around it throughout much of the trajectory. Lock-and-key docking on the curved wall is highly reliable. We attribute this robust interaction to gradients in the director field present everywhere near the curved wall that guide particles toward dales, and repel them from other locations on the wall, generating a self-correcting assembly mechanism. In comparison, sharp pyramidal docking sites,\cite{silvestre2014towards} allow particles to mis-assemble along their planar facets.

~\\

We were also able to probe different defect structures around particles. As mentioned, the relative size of the colloids to the LC cell thickness determines whether they form Saturn rings or elastic dipoles. In our experiments,  small variations in the size of the particles can favor dipole formation; for particles  of diameter $2a$ = 5 $\mu$m, the elastic dipole is the stable configuration. The elastic dipole also interacts with the curved wall, resulting in a different mode of attraction. The dipole can attach to either hills or dales, according to its orientation with respect to the wall (Fig. 4). Our data indicate that the interaction in this case acts at longer range, when the particles are a few tens of microns away from the features. The dipoles migrate towards the surface as shown in ESI Video 3 and 4 and, since their near field is asymmetrical, they dock selectively onto sites most compatible with the director field.  In the case of a dipole with the defect pointing away from the wall, the dale is still the preferred location (Fig. 4(a,b)), but if the defect is oriented towards the wall, the equilibrium position of the colloid is on the hill (Fig. 4(c,d)). The relative size of the colloid and the cavity also appears to be a crucial factor: small colloids ($2a$ = 5 $\mu$m) with a defect pointing away from the wall find their equilibrium location in the dales (this mechanism is robust: it occurs 90$\%$ of the time, i.e. in 58 out of 64 observations), while in large colloids with dipolar defect ($2a$ = 12 to 15 $\mu$m) this is not observed.  

~\\

In discussing the range of interaction there are two relevant length scales we must consider, the wavelength of the structure and the extrapolation length. In small-slope approximation, assuming infinite anchoring, the disturbance created by the structure decays as $2 \pi \lambda^{-1}$ with distance from the structure. However, in real systems, anchoring energies can be finite, causing a second length scale to enter, i.e. the extrapolation length, the ratio of elastic constant and anchoring strength, typically also on the microscale. An exploration of system behavior for wavelength comparable to the extrapolation length would be an interesting issue to study in future work to explore how effects associated with finite anchoring energies weaken the elastic energy gradients. Finally, colloids that excite dipolar defects interact with the elastic energy field at the wall at greater distances than do those that excite quadrupolar defects, consistent with the notion that quadrupolar interactions decay faster than dipolar interactions.\cite{lubensky1998topological} 

\begin{figure}[h]
\centering
  \includegraphics[height=6.5cm]{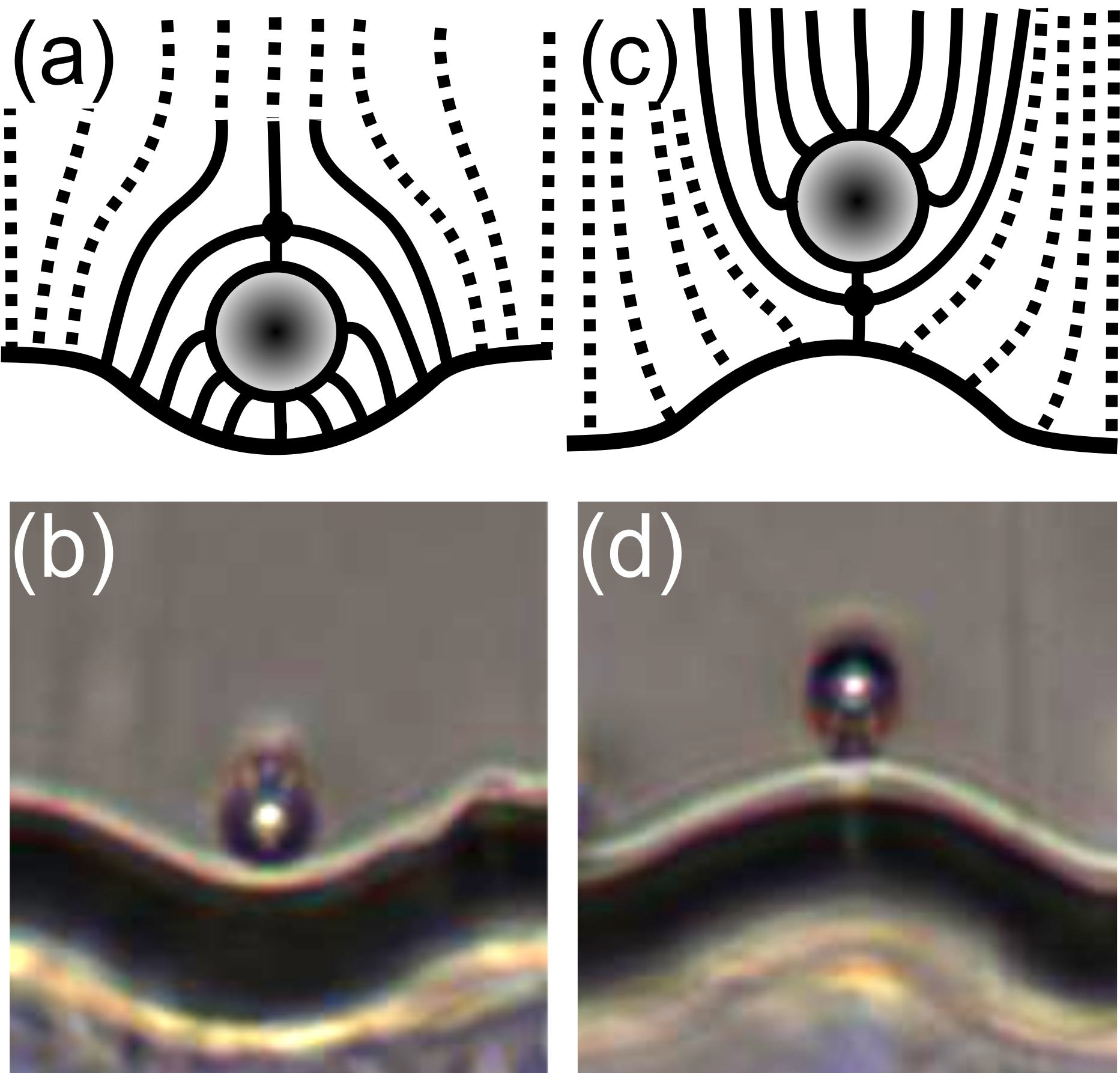}
  \caption{Schematic and bright field microscopy images of a particle ($2a$ = 5 $\mu$m) acting as (a,b) a dipole attracted to a dale with its point defect oriented outwards and (c,d) a dipole attracted to a peak with its point defect oriented towards the wall.}
\end{figure}

\begin{figure}[!h]
\centering
  \includegraphics[height=5cm]{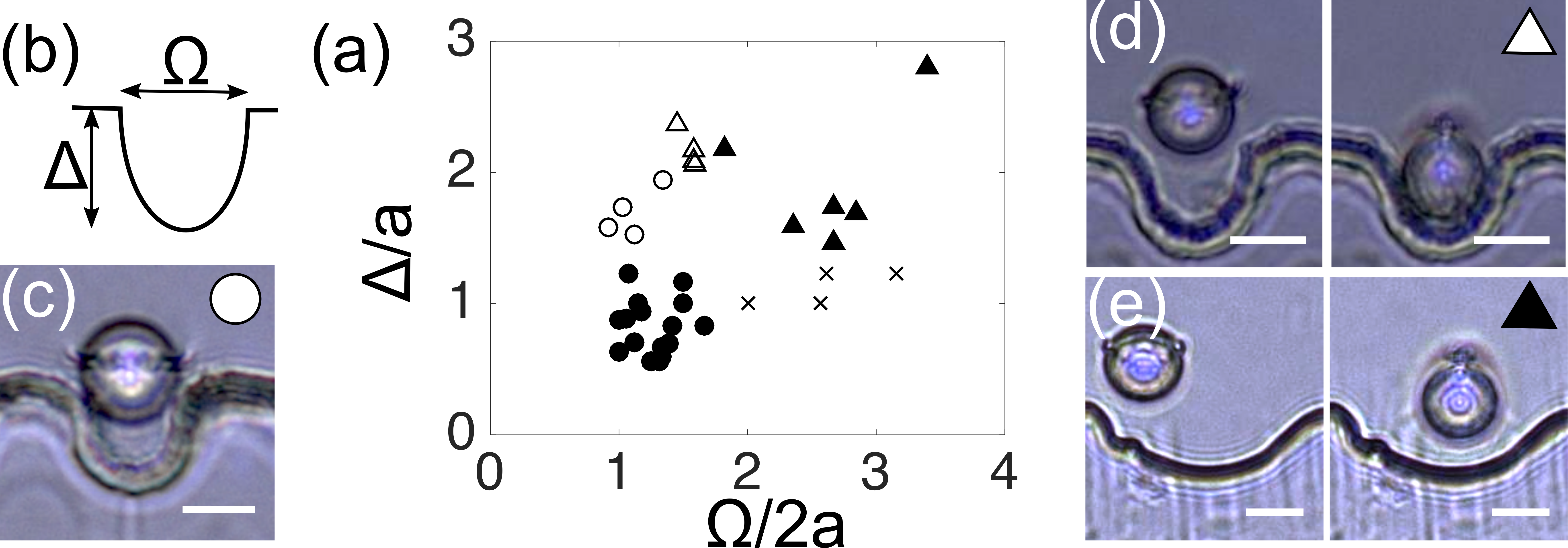}
  \caption{Colloids in wells of different dimensions. (a) Phase diagram of behaviors of colloids of radius $a$ in wells of width $\Omega$ and depth $\Delta$. Filled circles are cases of successful lock-and-key. Open circles are Saturn ring stuck on top of the wells. Triangles are quadrupole to dipole transitions which are stuck on top of the wells (filled) or enter the well (open). Crosses are cases where the colloids cannot dock. (b) Schematic of wells. (c) Particle with stable Saturn ring trapped at well entrance for $\Omega \approx 2a$ and $\Delta > a$. (d) Quadrupole to dipole transition for  $\Omega > 2a$; colloid docks with incomplete entry into well. (e) Quadrupole to dipole transition in a shallow well as the colloid finds an equilibrium distance from the wall. The scale bars are 10 $\mu m$.}
  \label{fig1}
\end{figure}


It is interesting to probe the colloid behavior as the  geometry of the ``lock'' is varied. We find a wide range of behaviors, summarized in Fig. 5(a), as we change the boundary from one that imposes dales, or shallow wells, to one with wells of greater depth, shown schematically in Fig. 5(b). We vary colloid diameter $2a$ over the range of 12-17 $\mu$m, well depths $\Delta$ from 4.5 to 21 $\mu$m, and well widths $\Omega$ from 16 to 51 $\mu$m. Our results indicate that confinement in both the lateral and vertical directions play important roles. When the well width is similar to the particle diameter, and its depth is similar to the particle radius, the lock and key mechanism occurs (filled dots). However, if the colloid is too wide (open dots) the particle is attracted toward the well but becomes stuck on top of it (Fig. 5(c)). Sometimes, in this case, the quadrupole can turn into a dipole pointing away from the wall (open triangles, Fig. 5(d)). If the cavity is too wide and not very deep, the particle does not dock (crosses). However, if the cavity is wide and sufficiently deep to fit the whole colloid inside, the particles does enter the cavity, and the Saturn ring switches to a hedgehog (filled triangles, Fig. 5(e)). Quadrupole-to-dipole transition has been previously induced by electrical field, \cite{loudet2001application} by magnetic field,\cite{stark2002saturn} and by flow.\cite{khullar2007dynamic} Here we show that a distorted curvature field, without external stimuli, can have a similar effect. These results suggest that, by tuning well shape, one can preserve the Saturn ring defect or eliminate it, an outcome of increased interest now with the advent of material assembly within such sites. \cite{wang2016experimental}

~\\


In conclusion, a new experimental system has been designed to study the interaction of colloidal particles with curved walls that act as sources of splay and bend distortions. The particles migrate towards sites with complementary geometry and similar anchoring. The ability to turn repulsion into attraction through the curvature of the wall is useful in providing templates to guide colloidal assembly. The distortion of the Saturn rings hints at the complex interaction between the particle's defect and the distorted director field around a structured wall. The range of interaction between the particles and the wall depends not only on the anchoring of liquid crystals on the particle, but also on the specific type of topographical feature of the wall, suggesting a possible route to sort particles with very small differences in sizes or anchoring strength. In the future, this versatile experimental system can be used to investigate the interaction between different types of ``locks'' and ``keys''. For example, the current experiments only considers spherical particles. However, the investigation could be easily expanded for study particles with different curvatures and anisotropic shapes. Ellipsoids, or faceted platelets, create distortions in the director field and defects which are different from the spherical particles,\cite{lapointe2009shape} therefore new types of locks, including faceted locks or those that orient high aspect ratio particles, might be carefully designed in terms of shapes and anchoring to have tunable attractions and selectivity of particles. The wall as a source of bend and splay could be recast in other geometries, including ``star-like'' cross sections that could seed structure growth in differing symmetries. This work could also be expanded to the design of particles with complementary shapes,\cite{sacanna2010lock} to promote or lock and key assembly in the bulk liquid crystal. This system could also be explored for sensing and detection purposes. An example of a sensor sensitive to anchoring was explored in the work by Lin, {\it et al.}\cite{abbott} for colloidal particles on a nematic droplet. 

~\\

With respect to the last two examples, the system presented here has the advantage that the location of the sorted particles can be controlled by placing the topographical features on desired locations on a 2D device. This can be exploited, for example, in the design of a microfluidic chip. By varying such conditions as the particles' size and anchoring, the wall anchoring and geometry, this system is a simple and effective playground to explore various schemes of directed self-assembly. We have shown in the past that sharp corners are able to attract particles at long-range \cite{cavallaro2013exploiting,luo2016around} and that the influence of the corners could be felt by particles tens of microns away. In this system, the range of interaction is much smaller. This suggests that ever more exquisite control of the directed assembly of colloids in NLCs can be achieved by combining topographical features that exert a long-range interaction with others that locally guide their position. \newline

~\\

We thank John C. Crocker, Randall D. Kamien, Shu Yang, Ali Mozaffari, Nima Sharifi-Mood, and Nuno M. Silvestre for useful discussions. This work was supported by the National Science Foundation (NSF) Materials Science and Engineering Center (MRSEC) Grant to the University of Pennsylvania, DMR-1120901. 

\begin{appendices}
\section{Drag coefficient analysis}

\par The average velocity v throughout the trajectory is 1 $\mu$m/s and the radius of the colloid a = 8 $\mu$m. Using typical material constants density $\rho$ = 1.38 $g/cm^3$, viscosity $\eta$ = 14.31 $mPa\cdot s$, elastic constant K = $10^{-12}$ $g/cm^3$, we can find $Re = \frac{\rho v a}{\eta}$ =  3 $\times 10^{-6} \ll 1$ and $Er =\frac{\rho v a}{K} = 0.1 \ll 1 $. Therefore the viscous force balances the elastic force and elastic forces will exceed the viscous forces and so the director field will not be strongly affected by the flow field.

~\\

Since the particle moves near the wall is in close proximity to three surfaces (those that form the top and bottom of the cell, and the wavy wall), corrections to the viscous drag on the sphere must be addressed to account for interactions with these boundaries. Since there are no exact analyses or simulations that address this geometry, we approximate the drag coefficient as the sum of two contributions, $C_{gap}$ accounting for the sphere in a thin gap, using the analysis of Ganatos \cite{ganatos1980strong}, the other, $C_{\parallel}$ accounting for the motion of a sphere parallel to a bounding wall, adopting the results of O'Neill \cite{o1964slow}. In addition, the particle moves toward the wavy wall, which could require an additional contribution to the drag coefficient, $C_{\bot}$. We estimate the magnitude of this contribution using the analysis of Brenner \cite{brenner1961slow}, and find that the changes in the drag coefficient for the wall distances explored by our particles is sufficiently weak that this contribution can be ignored.

~\\

The motion with respect to the wavy wall can be decomposed into $v_{\bot}$ and $v_{\parallel}$:

\begin{equation}
    \left\{
                \begin{array}{ll}
                  v = \sqrt{v_\bot^2-v_\parallel^2} \\
                	v_\bot = vcos(\theta - \psi)\\
		v_\parallel= vsin(\theta + \psi) \\
		                \end{array}
              \right.
\end{equation}

\noindent where $\theta = tan^{-1}\Big(\frac{v_y}{vx}\Big)$ and $\psi= tan^{-1}\Big(\frac{dy}{dx}\Big) = tan^{-1}(h'(x))$,  assuming the wavy surface is located at $y = h(x)$.

~\\

Brenner \cite{brenner1961slow} studies the interaction of a spherical moving perpendicular to a surface, where the Stokes' drag is corrected by a factor of $C_{\bot}$ as a function of normalized distance to the wall $\alpha = \frac{d}{a}$:

\begin{equation}
\frac{F_{\bot}}{6\pi\eta a v_{\bot}} = C_{\bot} (\alpha)
\end{equation}\

$C_{\bot}$ is calculated via Equation (2.19) in Brenner's paper, the drag correction values are tabulated in Table 1 in the above reference. At $\alpha = 1$, $C_{\bot} \to \infty$, but for the course of our trajectory, $C_{\bot}$ varies from 1.6 to 1.9.   
\par
The parallel motion of a sphere near the wall is studied by O'Neill \cite{o1964slow}, the correction to Stokes' equation owing to the presence of the wall is:

\begin{equation}
\frac{F_{\parallel}}{6\pi\eta a v_{\parallel}} = C_{\parallel} (\alpha)
\end{equation}

\noindent where $C_{\parallel}$ is also a function of $\alpha$. It is determined by Equation (26) and tabulated in \cite{o1964slow}.

Based on the $\alpha$ from our trajectory, $C_{\bot}$ and $C_{\parallel}$ both vary from 1.3-1.9 (as shown in Fig. 1). 

 \begin{figure}[h]
\centering
  \includegraphics[height=6cm]{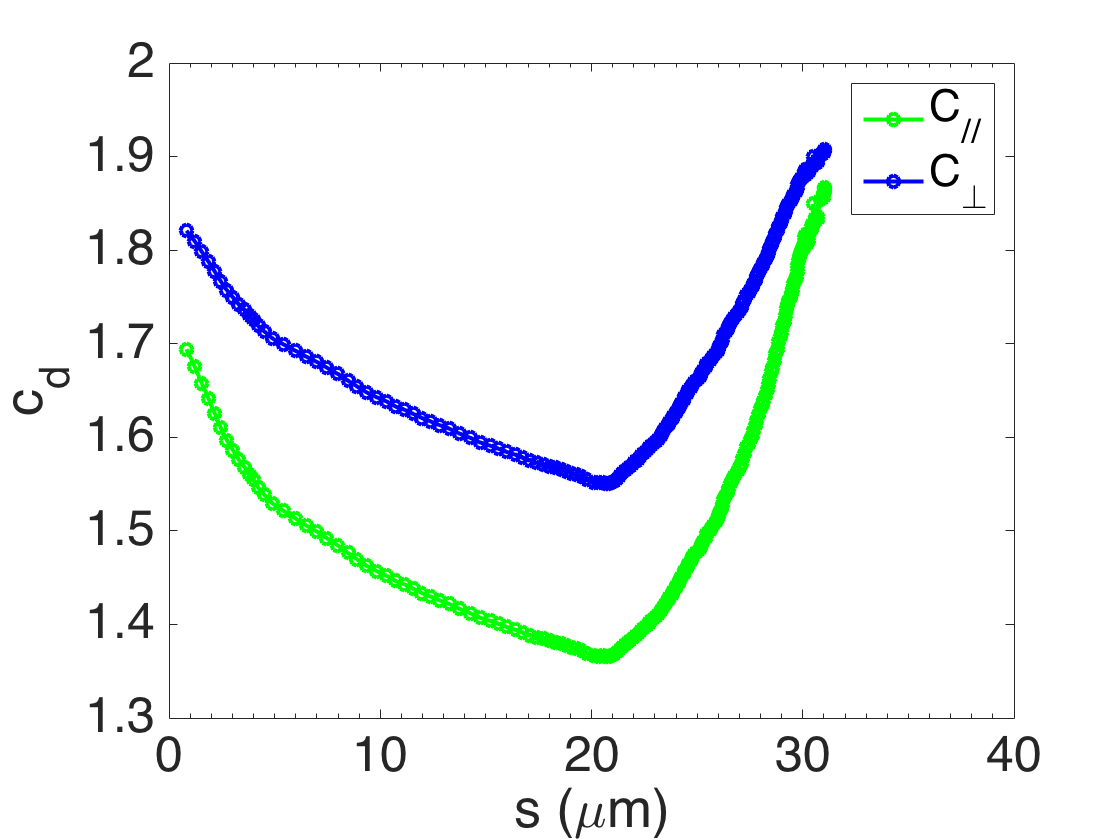}
  \caption{Drag coefficients at each arclength s from Brenner ($C_{\bot}$) and O'Neill ($C_{\parallel}$) calculated from the distance of the colloid from the wall at each point along the trajectory}
  \label{fig1}
\end{figure}

The movement is also constrained in $\hat z$.  Given the symmetries of the experimental configuration and the elastic repulsion from the gap walls, it is reasonable to assume that the particle is centered in the gap. The drag coefficient  is given graphically in Fig. 3(a) in terms of $\beta = \frac{l}{a}$  for a gap of half-width $l$ comparable to the sphere radius $a$ in the analysis of Ganatos, Pffefer and Weinbaum \cite{ganatos1980strong}.  From that graph, for spheres and gaps of size that correspond to our experiment, the combined effect of two surfaces yields $C_{gap}=3$. 

~\\

Furthermore, the anisotropy of viscosities must be accounted for in LC. For micron-sized particle moving in 5CB, typical values of viscosities are $\eta_{\bot} = 14.3$ $mPa\cdot s$ and $\eta_{\parallel} = 7.83$ $mPa\cdot s$. Our trajectories consist of prolonged motion parallel to the wall (transverse to the director field), followed by a reversal and docking event (parallel to the director field, with significant near field splay-matching interactions. For this docking segment, the more relevant viscosity may be $\eta_{\parallel}$. 

~\\

Our aim is to extract an order of magnitude estimate for the interaction energy.  We do this in two ways. In one, we use the transverse viscosity $\eta_{\bot}$ for the entire trajectory, and estimate the net drag coefficient from the gap and interactions with the wavy wall to be about $C_D \approx 4$. Based on this correction factor, the energy $U$ can be found by integrating the velocity over the entire trajectory:

\begin{equation}
U = {C_D} 6\pi \eta a\int\limits_0^{{s_f}} {v} d{s}
\end{equation}

By this method, we find $U \approx 10^5$ $k_BT$. 

~\\

In a  more careful treatment, we decompose the motion into $v_\bot$ and $v_\parallel$. We assume the colloid is close enough to the surface that the relevant viscosity for $v_\bot$ is $\eta_\parallel$, and that for $v_\parallel$ is $\eta_\bot$. We let $C_{gap} \approx 3$. The drag force can be decomposed as follows:

 \begin{equation}
    \left\{
                \begin{array}{ll}
                 F_{\parallel, gap} = C_{gap} 6\pi{\eta_{\bot}} a v \\
		F_{\parallel, wall} = C_{\parallel}(\alpha) 6\pi{\eta_\bot} a v_\parallel \\
		F_{\bot, wall} =C_{\bot}(\alpha) 6\pi{\eta_\parallel} a v_\bot
                \end{array}
              \right.
\end{equation}

The energy dissipation throughout the trajectory, summing up all contributions:

\begin{equation}
U = 6\pi a\left( {{\eta _{\parallel}}\int\limits_0^{{s_f}} {{C_{gap}}} {v} d{s} + {\eta _{\parallel}}\int\limits_0^{{s_f}} {{C_{\parallel}}} {{v}_{\parallel}} d{{s}_{\parallel}} + {\eta _{\bot}}\int\limits_0^{{s_f}} {{C_{\bot}}} {{v}_{\bot}} d{{s}_{\bot}}} \right)
\end{equation}

\noindent where ${s}_{\parallel}$ and ${s}_{\bot}$ are the parallel and perpendicular components of the arc length s. ${s}_{\parallel} = s\cdot sin(\theta - \psi)$ and ${s}_{\bot} = s\cdot cos(\theta - \psi)$.

~\\

By this method, we find $U$ to differ by 2 percent from the  one viscosity and one drag coefficient approximation. This small difference is owing to the fact that the rate of migration for the last stage of docking after the reversal is extremely slow. For the purposes of an order of magnitude estimate, we give the simpler discussion in the main text. 

\end{appendices}

\bibliographystyle{unsrt}

\bibliography{rsc}

\end{document}